\documentclass[pra,showpacs,twocolumn,superscriptaddress,amsmath,amssymb]{revtex4-1}
\usepackage{bm}
\usepackage{graphicx}
\usepackage{color}
\usepackage{dcolumn,amsmath}
\usepackage[normalem]{ulem}
\usepackage[version=4]{mhchem}

\begin{document}

\title{Long-lived complexes and signatures of chaos in ultracold K$_2$+Rb collisions}

\author{J. F. E. Croft}
\author{N. Balakrishnan}
\affiliation{Department of Chemistry, University of Nevada, Las Vegas, Nevada 89154, USA}
\author{B. K. Kendrick}
\affiliation{Theoretical Division (T-1, MS B221), Los Alamos National Laboratory, Los Alamos,
New Mexico 87545, USA}

\begin{abstract}
Lifetimes of complexes formed during ultracold collisions are of current
experimental interest as a possible cause of trap loss in ultracold gases
of alkali-dimers.
Microsecond lifetimes for complexes formed during ultracold elastic collisions of K$_2$
with Rb are reported, from numerically-exact quantum-scattering calculations.
The reported lifetimes are compared with those calculated using a simple density-of-states
approach, which are shown to be reasonable.
Long-lived complexes correspond to narrow scattering resonances which we examine
for the statistical signatures of quantum chaos, finding that the positions
and widths of the resonances follow the Wigner-Dyson and Porter-Thomas
distributions respectively.
\end{abstract}
\maketitle

\section{Introduction}
The ability to create ultracold atomic gases has engendered extraordinary
progress in a diverse range of physical areas.
In few-body physics the first experimental observation of Efimov states was
made in a gas of ultracold ceasium~\cite{kraemer.mark.ea:evidence};
in many-body physics experimental observation of zero-temperature quantum phase
transitions have been made in a gas of ultracold rubidium atoms trapped in an
optical lattice~\cite{greiner.mandel.ea:quantum};
in precision measurement the development of new atomic-clock technologies
has used ultracold strontium trapped in an optical lattice~\cite{bloom.nicholson.ea:optical}.
What all these diverse applications have in common is that they take
advantage of the exquisite precision and control uniquely attainable in the
ultracold regime.

Ultracold molecules share the precision and control of ultracold
atomic gases, while their richer internal structure and long-range anisotropic
interactions open up an even more diverse range of applications.
Cold and ultracold molecules have been used to study chemical reactions
at their most fundamental level~\cite{ospelkaus.ni.ea:quantum-state},
measure the shape of the electron~\cite{hudson.kara.ea:improved,baron.campbell.ea:order},
and perform precision spectroscopy of complex molecules~\cite{spaun.changala.ea:continuous}.
There is a wide array of theory proposals for ultracold molecular samples,
which take advantage of the complexity of molecules relative to atoms,
from studying novel quantum phases~\cite{micheli.pupillo.ea:cold,wall.carr:emergent,lechner.buchler.ea:role,ewart.wall.ea:bosonic}
to quantum information processing~\cite{buchler.demler.ea:strongly,andre.demille.ea:coherent,yelin.kirby.ea:schemes,demille:quantum}.
Such proposals rely on the ability to produce a stable ultracold molecular
gas without significant loss.

The rich structure of molecules is however both a blessing and a curse.
Trap loss has been found to be a limiting factor in ultracold molecule experiments
with alkali dimers
\cite{takekoshi.reichsollner.ea:ultracold,molony.gregory.ea:creation,park.will.ea:ultracold,guo.zhu.ea:creation}.
This loss persists even when the molecules are in their absolute
ground-state and only elastic collisions are possible.
One possible mechanism for this loss is the formation of long-lived complexes.
Based on statistical arguments, Mayle {\it et al\/} have proposed that due to the
high density-of-states (DOS) in such systems 4-body complexes can have lifetimes
of order 1-–10~ms~\cite{mayle.ruzic.ea:statistical, mayle.quemener.ea:scattering}.
If these lifetimes are reasonable then this could explain the experimentally
observed trap loss, which limits trap lifetime to the order of seconds.
Lifetime estimates for 3-body alkali complexes formed in atom-dimer collisions
using this DOS approach have been shown to be reasonable when compared to
estimates from classical trajectory calculations~\cite{croft.bohn:long-lived}.
So far, however, there has been no explicit experimental measurement or theory
predication based on quantum calculations for such lifetimes.

Such complexes can however also be considered a feature not a bug.
They are expected to exhibit the Wigner-Dyson energy level statistics
associated with quantum chaos.
There has been recent interest in understanding the role of chaos in cold collisions
\cite{bohn.avdeenkov.ea:rotational,flambaum.ginges:resonance,croft.bohn:long-lived,frye.morita.ea:approach,
croft.makrides.ea:universality,green.vaillant.ea:quantum,frisch.mark.ea:quantum,maier.kadau.ea:emergence,
mayle.ruzic.ea:statistical, mayle.quemener.ea:scattering,maier.ferrier-barbut.ea:broad,
yang.perez-ros.ea:classical,jachymski.julienne:chaotic}.
Ultracold molecular collisions are complex and understanding them in
chaotic terms could allow for the considerable insight gained in nuclear
physics to be applied to ultracold atomic and molecular physics~\cite{weidenmuller.mitchell:random}.
For example Mayle {\it et al\/} have developed a statistical scattering formalism which
assumes the formation of chaotic complexes at short range
\cite{mayle.ruzic.ea:statistical, mayle.quemener.ea:scattering}.

In this work we report explicit delay times for elastic K$_2$-Rb collisions,
obtained from numerically-exact quantum-scattering calculations.
We compare these delay times with lifetimes for the collision complex predicated
by a simple DOS approach finding them to be reasonable and validating their use for other
similar systems.
We also examine the statistics of resonance positions and widths finding that they
follow the Wigner-Dyson and Porter-Thomas distributions respectively.
Both distributions are characteristic of quantum chaos.

\section{Methods}
\begin{figure}[tb]
\centering
\includegraphics[width=1\columnwidth]{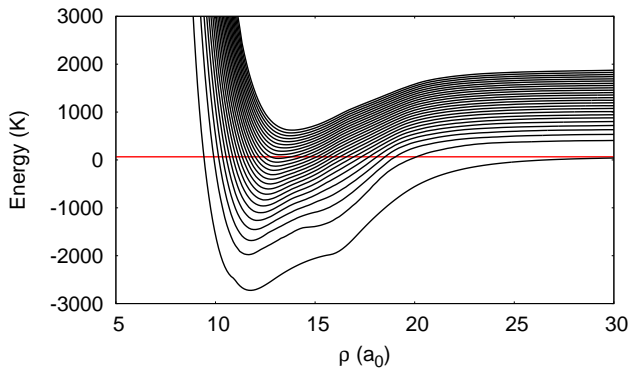}
\caption{The adiabatic potential curves for the KRbK complex at short hyperradius
for even exchange symmetry.
Only every 100th curve is shown due to the high DOS for this system. The horizontal
red line corresponds to the $v=0$ $j=0$ threshold of K$_2$.}
\label{fig:aphsf}
\end{figure}

We use the atom-diatom scattering formalism as developed by Pack and Parker
\cite{pack.parker:quantum,kendrick.pack.ea:hyperspherical}.
In the short range we use adiabatically-adjusting principle-axis hyperspherical
(APH) coordinates, an approach which ensures that all arrangements are treated fully
equivalently, while in the long range we use Delves hyperspherical coordinates.
The wavefunction in the short-range APH region is expanded in an orthonormal
basis capable of accurately representing the hyperspherical harmonics
\cite{kendrick.pack.ea:hyperspherical}.
This basis is therefore capable of fully representing the complex collision
dynamics of 3 atoms at short range which is the source of the
long lifetimes predicted by Mayle {\it et al}
\cite{mayle.ruzic.ea:statistical, mayle.quemener.ea:scattering}.
We emphasize that while in this work we are primarily interested in elastic scattering
this reactive formalism is in fact required as many of the channels open at
short hyper-radius correspond asymptotically to configurations with a bound
KRb dimer.
The lowest 2500 adiabatic potential curves are shown in figure~\ref{fig:aphsf} where
the high DOS which leads to the complex short-range dynamics can
easily be seen.
All calculations are for total angular momentum $J=0$ even parity and even
identical-particle exchange symmetry.
Coupling of the orbital angular momenta with both the electron and nuclear spins
is omitted.
For non-zero $J$, the computational cost is prohibitive scaling as
$\mathcal{O}((J+1)^3)$, even when we take advantage of parity and exchange symmetries.
Fortunately, we are primarily interested in the  ultracold regime where only
$s$-wave collisions contribute (that is, only $J = 0$ is required for K$_2$ in
the ground rotational state $j = 0$).
The log-derivative matrix was propagated using the method of Johnson~\cite{johnson:multichannel}.
The details of the calculation are the same as used for the
K+KRb reaction, for details see~\cite{croft.makrides.ea:universality}.
An {\it ab initio} ground-state potential energy surface was used
which accurately accounts for the long-range dispersion behavior,
for more details see~\cite{croft.makrides.ea:universality}.

\section{Results}
\subsection{Long-lived complexes}
The long lifetimes predicted for complexes in ultracold alkali-dimer collisions
are due to two main factors.
Firstly deep potentials and heavy atoms lead to a high DOS,
classically this corresponds to many ro-vibrational degrees-of-freedom for the
energy to distribute into.
Secondly few exit channels mean the complex spends a long time exploring these
degrees-of-freedom before finding a way out.
These concepts are codified by Rice-Ramsperger-Kassel-Marcus
(RRKM) theory~\cite{marcus:lifetimes,marcus:lifetimes*1,levine:molecular}.
RRKM theory predicts a lifetime given by,
\begin{equation}
\tau = \frac{2 \pi \hbar \rho}{N_\mathrm{o}}.
\label{eq:rrkm}
\end{equation}
Where $\rho$ is the DOS and $N_\mathrm{o}$ is the number of energetically allowed
exit channels.
The RRKM lifetime originated in transition state theory however it has been
used by Mayle {\it et al} as a way to estimate lifetimes in ultracold molecular
collisions~\cite{mayle.ruzic.ea:statistical,mayle.quemener.ea:scattering}.
The beauty of equation~\ref{eq:rrkm} is its simplicity,
only an estimate of the DOS and the number of open channels is needed to
calculate a complex lifetime at a given energy.

We proceed to calculate the lifetime of a KRbK complex, formed in elastic K$_2$+Rb
collisions, by following the method detailed by Mayle {\it et al} for estimating
$\rho$~\cite{mayle.ruzic.ea:statistical}.
For the K$_2$ dimer potential we use a Lennard-Jones potential with
$C_6$ and $D_e$ taken from~\cite{falke.knockel.ea:potassium}.
To obtain the $C_6$ and $D_e$ required for the Lennard-Jones K$_2$-Rb potential
we use the $C_6$ and $D_e$ for KRb taken from~\cite{pashov.docenko.ea:coupling}
and assume the three-body potential is pairwise additive, with C$_6$ and $D_e$
chosen to be double the atom + atom value for the atom + dimer potential.
In this way the choice of $C_6$ used to estimate the DOS is the same as the
$C_6$ used in the scattering calculations.
The 1d-Schr\"{o}dinger equation was solved using the
Fourier-grid-Hamiltonian method~\cite{balint-kurti.dixon.ea:grid,marston.balint-kurti:fourier}.
Only even rotational levels are included for K$_2$ to account for the identical
particle symmetry.
Using this approach the estimate for $\rho$ is 3.5~mK$^{-1}$ which gives a
complex lifetime of 167~ns.

\begin{figure}[tb]
\centering
\includegraphics[width=1\columnwidth]{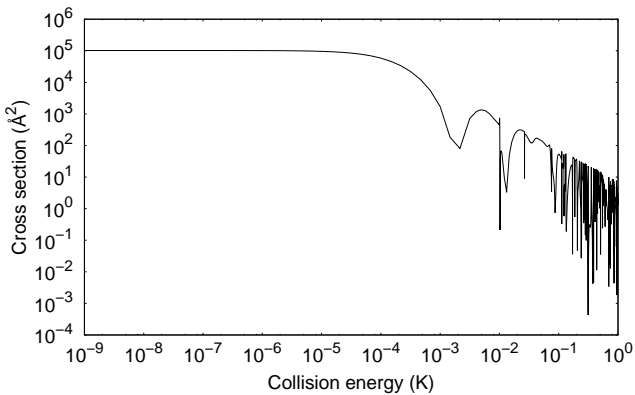}
\caption{Elastic cross-section for collisions of K$_2$($v=0,j=0$) with Rb as
a function of energy, for total angular momentum $J=0$.}
\label{fig:cross_section_log}
\end{figure}

The long lifetimes based on the DOS approach manifest themselves quantum
mechanically as scattering resonances.
This can be understood intuitively from the energy-time uncertainty principle,
\begin{equation}
\Delta E \Delta t \ge \frac{\hbar}{2}.
\label{eq:uncertainty}
\end{equation}
This can be interpreted as saying that a narrow resonance implies a long
lifetime,
\begin{equation}
\tau \approx \frac{\hbar}{2\Gamma},
\label{eq:lifetime}
\end{equation}
where $\Gamma$ is the width of the resonance and $\tau$ is the
lifetime, which quantify the uncertainty in energy and time respectively.
One has to be careful interpreting these equations as time is not an
observable in quantum mechanics, however they do make clear that long-lived
complexes  correspond to narrow resonances in quantum scattering.
Figure~\ref{fig:cross_section_log} shows the elastic cross-section for
collisions of K$_2$($v=0,j=0$) with Rb as a function of energy,
for total angular momentum $J=0$.
It is seen that there is a forest of narrow resonances starting at about 10~mK,
each of which corresponds to a long-lived KRbK complex.
The zero-energy cross-section gives a scattering length of 90~\AA\ which
compares well to the average scattering length of 47~\AA\ obtained
from the $C_6$ coefficient~\cite{gribakin.flambaum:calculation}.

While equations~\ref{eq:uncertainty} and \ref{eq:lifetime} give an intuitive
understanding of why narrow resonances correspond to long lifetimes, we can be
more rigorous by computing Smith's $\mathbf{Q}$ matrix,
\begin{equation}
\mathbf{Q} = i\hbar \mathbf{S} \frac{\partial \mathbf{S}^{\dagger}}{\partial E}.
\end{equation}
The eigenvalues of $\mathbf{Q}$ are the time delays between a collision with
and without a potential~\cite{smith:lifetime,smith:chapter}.
The trace of $\mathbf{Q}$ is therefore the sum of the delay times for each channel.
The matrix $\mathbf{Q}$ can be obtained directly by propagating the energy-derivative
of the log-derivative matrix, which can be done efficiently at the same time
as the log-derivative propagation~\cite{walker.hayes:direct}.
This allows for the direct calculation of $\mathbf{Q}$ from the energy
derivatives of $\mathbf{K}$, and $\mathbf{S}$.
This approach has been extensively used to compute smith delay times in a
variety of systems and contexts
\cite{kendrick.pack:recombination,kendrick.pack:geometric*2,guillon.stoecklin:analytical}.

\begin{figure*}[tb]
\centering
\includegraphics[width=1\textwidth]{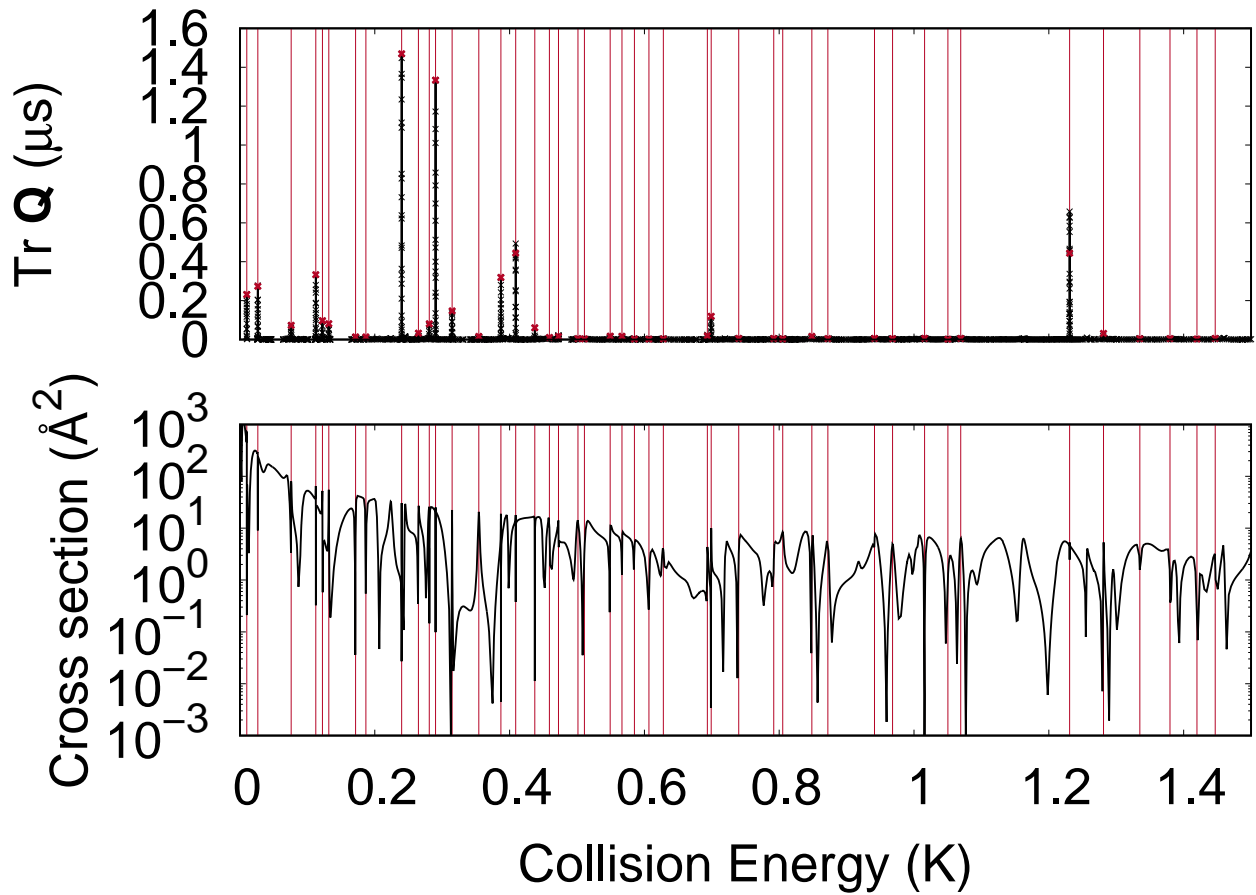}
\caption{Upper panel: $\mathrm{Tr}(\mathbf{Q})$ as as a function of collision energy.
Lower panel: elastic cross-section for collisions of K$_2$($v=0,j=0$) with Rb as
a function of energy.
The vertical red lines mark the resonance positions while the red crosses show
the resonance lifetimes.
The K$_2$ $v=0, j=2$ channel becomes open at 0.47~K.}
\label{fig:life_cs}
\end{figure*}

Figure~\ref{fig:life_cs} plots $\mathrm{Tr}(\mathbf{Q})$ as a function of energy
along with the cross-section on the same energy grid.
The lifetimes are computed on a energy grid with over 1600 points. The small gaps
seen in the energy grid are regions where the delay time is negative.
Smith time delays are the difference between the time for a collision and the time
for a collision without a potential, we refer to the latter as the background
collision time.
This definition leads to negative delay times away from resonances as the
background collision time is large due to the lack of an attractive potential,
compared to the collision time with a potential, especially at low collision
energies.
It is easily seen that there are many resonances which exceed the DOS estimate
for the lifetimes of 167~ns.

Following \cite{kendrick.pack:geometric*2} we fit each of the resonances to
a Breit-Wigner form,
\begin{equation}
Q(E) = 2\hbar \frac{\Gamma_\mathrm{r}/2}{[(E_\mathrm{r}-E)^2+(\Gamma_\mathrm{r}/2)^2]}.
\end{equation}
Where $E_\mathrm{r}$ and $\Gamma_\mathrm{r}$ are the resonance energy and width
respectively.
This allows us to unambiguously assign a delay time,
\begin{equation}
Q_\mathrm{r}=\frac{4\hbar}{\Gamma_\mathrm{r}},
\end{equation}
to each resonance seen in figure~\ref{fig:life_cs}.
The parameters $E_\mathrm{r}$ and $Q_\mathrm{r}$ are shown on figure~\ref{fig:life_cs} as
solid red vertical lines and red dots respectively.
The delay times for the 10 narrowest resonances are given in Table~\ref{tab:res}.
Having assigned to each resonance an explicit delay time we can compare them
to the DOS estimate of 167~ns. We find that many resonances have lifetimes
longer than the DOS estimate.
In fact the lifetimes of the narrowest resonances are around an order of
magnitude longer than that predicted by the DOS approach.
It is noticeable however that we find far fewer such
resonances than we would expect based on the DOS used to estimate the lifetime.

The delay time is a well defined quantity however it does not correspond to the
lifetime as experimentally understood.
As discussed earlier the Smith delay time includes a significant
negative contribution from the collision partners traversing the long range
which does not correspond to what would generally be considered the lifetime of a collision complex
(the time collision partners spend strongly interacting at ``short range''
before eventually escaping to infinity).
Due to this difference the delay time should therefore be considered a lower
bound on the complex lifetime of experimental relevance.

The results shown here suggest that the DOS lifetime estimates for complexes
formed during ultracold collisions between molecules similar to those
studied here are also reasonable.
This in turn suggests that long-lived 4-body complexes formed in ultracold
dimer-dimer collisions are the cause of trap loss seen experimentally.

\subsection{Quantum chaos}
Quantum-scattering calculations for collisions of ultracold molecules are
extremely computationally expensive.
Even though here we have presented results for only total angular momentum $J=0$,
even exchange symmetry and parity, included no spin or field effects the
calculations still required over 300,000 hours of CPU time.
It is possible that numerically exact quantum scattering calculations including
all the effects omitted in these calculation will never be computationally
tractable.
As such statistical approaches offer an alternative way to attack
such problems~\cite{flambaum.ginges:resonance,mayle.ruzic.ea:statistical,mayle.quemener.ea:scattering,croft.bohn:non-sticking}.

Atom-dimer ultracold collisions have been shown to be classically chaotic~\cite{croft.bohn:long-lived}
and the analysis of short range adiabats of KRbK have been shown to
exhibit the characteristics of quantum chaos~\cite{croft.makrides.ea:universality}.
While quantum systems cannot exhibit the non-linearity characteristic of
classical chaos (in quantum mechanics operators are linear~\cite{von-neumann:mathematical})
quantum analogs of classically chaotic systems do exhibit certain statistical
signatures~\cite{bohigas.giannoni.ea:characterization}.
Such as Wigner-Dyson energy-level
statistics~\cite{wigner:on*1,dyson:statistical},
Ericson fluctuations~\cite{ericson:fluctuations,ericson:theory},
and Porter-Thomas resonance-width statistics~\cite{porter.thomas:fluctuations}.

\begin{figure}[tb]
\centering
\includegraphics[width=1\columnwidth]{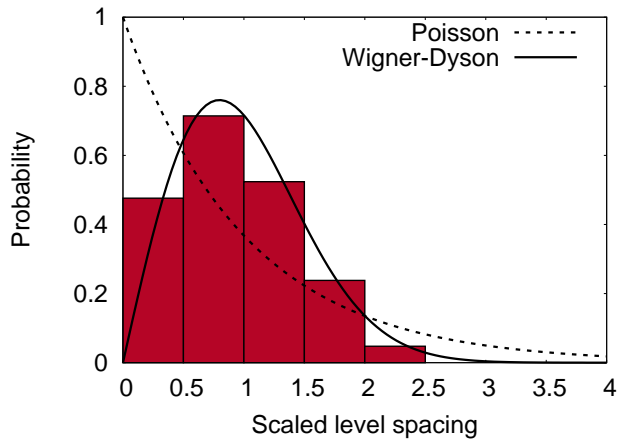}
\caption{Distribution of scaled nearest-neighbor spacings between scattering resonances,
shown as solid red lines in figure~\ref{fig:life_cs}.}
\label{fig:brody}
\end{figure}

We now proceed to examine the statistical distribution of resonance positions for evidence
of quantum chaos.
The distribution of scaled nearest-neighbor spacings for non-chaotic systems
is given by the Poisson distribution,
\begin{equation}
P_\mathrm{p}(s) = \exp{(-s)},
\end{equation}
where small spacings predominate.
However in quantum systems chaos manifests itself in the repulsion between energy
levels~\cite{wigner:on*1}, the distribution of scaled nearest-neighbor spacings is
then given by a Wigner-Dyson distribution~\cite{dyson:statistical}.
For Hamiltonians with time-reversal symmetry, such as we have here, the
nearest-neighbor spacings are given by,
\begin{equation}
P_\mathrm{wd}(s) = \frac{\pi}{2}s\exp{(-\frac{\pi}{4}s^2)}.
\end{equation}
Figure~\ref{fig:brody} shows the distribution of scaled nearest-neighbor spacings
between the 44 resonances shown in figure~\ref{fig:life_cs}.
While we have not found enough resonances to make a definitive statement
as to whether this system exhibits quantum chaos we do clearly see the
repulsion between neighbouring resonances characteristic of quantum chaos.

\begin{table}[tb]
\begin{tabular}{ccc}
\hline
   Energy~(K) &  Lifetime~($\mu$s) &  Width~($\mu$K) \\
\hline
     0.24 &       1.47 &   20.80 \\
     0.29 &       1.33 &   22.90 \\
     0.41 &       0.44 &   68.75 \\
     1.23 &       0.44 &   68.83 \\
     0.11 &       0.33 &   91.63 \\
     0.39 &       0.32 &   95.72 \\
     0.03 &       0.27 &  111.65 \\
     0.01 &       0.23 &  132.01 \\
     0.31 &       0.15 &  207.77 \\
     0.70 &       0.12 &  256.15 \\
\hline
\end{tabular}
\caption{Position, lifetime and width of the 10 narrowest resonances.}
\label{tab:res}
\end{table}

The degree to which the observed distribution matches the Wigner-Dyson
distribution can be quantified by the Brody parameter~\cite{brody:statistical,brody.flores.ea:random-matrix}.
The Brody parameter is itself not physically meaningful, rather is defined to
smoothly interpolate between the Poisson distribution and the Wigner-Dyson
distribution,
\begin{equation}
\begin{split}
P_\mathrm{b}(s) &= As^\eta\exp{(-\alpha s^{\eta+1})}, \\
A &= (\eta + 1)\alpha, \\
\alpha &= \Gamma\Big(\frac{\eta+2}{\eta+1}\Big)^{\eta+1},
\end{split}
\label{eq:brody}
\end{equation}
where $\eta$ is the Brody parameter.
For $\eta=0$ the distribution reduces to $P_\mathrm{p}$ and for $\eta=1$ it
reduces to $P_\mathrm{wd}$.
Performing a least-squares fit of equation~\ref{eq:brody} to the data shown in
figure~\ref{fig:brody} we obtain a Brody parameter of $\eta = 0.78 \pm 0.4$.
Despite the relatively small number of resonances found this value is clearly
suggestive of a chaotic system.

\begin{figure}[tb]
\centering
\includegraphics[width=1\columnwidth]{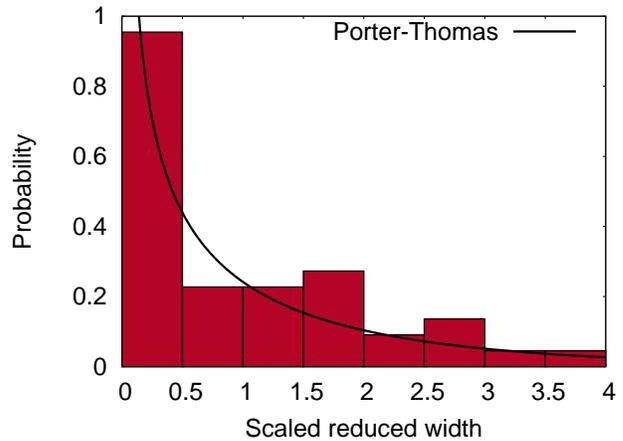}
\caption{Distribution of scaled reduced widths for scattering resonances,
shown as solid red lines in figure~\ref{fig:life_cs}.}
\label{fig:porter_thomas}
\end{figure}

We now move on to examine another statistical characteristic of chaos in quantum systems,
the Porter-Thomas distribution of resonance widths~\cite{porter.thomas:fluctuations}.
Porter-Thomas statistics describe the distribution of velocity-independent reduced widths,
\begin{equation}
\Gamma_n^0 = \Gamma_n / E_0^{\frac{1}{2}}.
\end{equation}
The reduced widths for chaotic systems follow the $\chi^2_k$ of degree $k=1$,
\begin{equation}
P_\mathrm{pt}(s) = s^{-\frac{1}{2}}e^{-\frac{1}{2}s}
\end{equation}
where the scaled reduced width is given by $s = \Gamma^0_n / \langle\Gamma^0_n\rangle$.
Figure \ref{fig:porter_thomas} shows the distribution of reduced widths for the
resonances shown in figure~\ref{fig:life_cs}.
Despite the relatively small statistical sample we clearly see broad agreement
with the overall trend of the Porter-Thomas distribution towards resonances
with smaller widths.

\section{Conclusions}
We have examined ultracold elastic K$_2$-Rb collisions and reported explicit
lifetimes for long-lived collision complexes formed during the collision process,
finding that such lifetimes can be of the order of microseconds.
These lifetimes were compared with those predicted using a simple DOS approach
based on RRKM theory which are shown to be reasonable.
The accuracy of such methods is of current interest as long-lived complexes
formed in ultracold dimer-dimer collisions have been proposed as the cause of
observed experimental trap loss.

Long-lived complexes correspond to narrow resonances in quantum scattering.
We have analyzed the distribution of nearest-neighbor spacings and widths of these
resonances and have found that they both exhibit the statistical signature
of quantum chaos.
Quantum scattering calculations for collisions of ultracold molecules are extremely
computationally expensive.
It is possible that numerically-exact quantum-scattering calculations including
all the effects of interest will never be computationally tractable.
As such the chaotic nature of system such as this suggests a statistical
approach to tackling such problems could be fruitful.

In future work we intend to examine the effect of including the excited doublet
state on the lifetimes. Including the excited state in the calculations will
allow for the full treatment of the conical intersection and the possibility of
long-lived quasi-bound states on the upper surface.

\section{Acknowledgments}
We acknowledge C. Makrides, M. Li, and S. Kotochigova for helpful
discussions.
We acknowledge support from the  US Army Research Office, MURI grant No.~W911NF-12-1-0476
(N.B.), the US National Science Foundation, grant No.~PHY-1505557 (N.B.).
BKK acknowledges that part of this
work was done under the auspices of the US Department of Energy,
Project No. 20170221ER of the Laboratory Directed Research and Development
Program at Los Alamos National Laboratory.
Los Alamos National Laboratory is operated by Los Alamos National Security,
LLC, for the National Security Administration of the US Department of Energy
under contract DE-AC52-06NA25396.
JFEC gratefully acknowledges support from the ITAMP visitor’s program.

\bibliography{../../all}

\end{document}